\documentclass[10pt]{iopart} \usepackage{graphicx,epsfig}

\begin{document}
\title[Deformation of dark solitons] {Deformation of dark solitons in inhomogeneous 
Bose-Einstein
condensates}

\author{N.G. Parker, N.P. Proukakis, M. Leadbeater, and C.S. Adams}
\address{Department of Physics, University of Durham, South
Road, Durham DH1 3LE, United Kingdom}
\ead{n.g.parker@durham.ac.uk}
\begin{abstract}
A dark soliton becomes unstable when it is incident on a background density gradient,
 and the induced instability results in the emission of sound.
Detailed quantitative studies of sound emission are performed for
 various potentials, such as steps, linear ramps and gaussian traps.
The amount of sound emission is found to be a significant fraction of the
soliton energy for typical potentials.
Continuous emission of sound is found to lead to an apparent deformation of
the soliton profile.
The power emitted by the soliton is shown to be parametrised by the
square of the displacement of the centre of mass of the soliton from its density minimum,
thus highlighting the significance of the inhomogeneity-induced soliton 
deformation.
 \end{abstract}
\maketitle

\section{Introduction}

The intrinsic
similarities between the Gross-Pitaevskii equation describing
the zero-temperature mean-field dynamics of Bose-Einstein condensates
(BEC) and the cubic nonlinear
Schrodinger equation (NLSE) of nonlinear optical systems implies
the existence of common fundamental phenomena, independent of the 
origin of the nonlinearity.  For example,
experiments on BEC's have recently led to the observation of
four-wave mixing of matter waves \cite{Deng}, and
both bright \cite{Strecker,Khaykovich} and dark \cite{Denschlag,Burger,Dutton} 
matter wave solitons.
Solitons are, strictly speaking, shape-preserving one-dimensional excitations
supported in nonlinear media, where the effect
of dispersion is balanced by the interatomic interactions.
Three-dimensional analogs of solitons,  so-called solitary waves,
do exist, but are more prone to instabilities than their one-dimensional
counterparts \cite{Kivshar2}.
In the context of nonlinear optics, solitons propagate in the
effectively one-dimensional (1D) medium of homogeneous optical fibers.
Bright solitons find applications in optical communications 
\cite{Agrawal}, 
whereas potential applications of the less
exploited dark solitons include the
formation of zero cross-talk optical junctions \cite{Miller}. 
In the case of atomic BEC's, 1D geometries have been recently engineered by
suitable combination of optical and magnetic traps \cite{Gorlitz,Schreck}, 
and also on microchips \cite{Hansel,Ott,Schneider}.

The one-dimensional Gross-Pitaevskii Equation (GPE) describing trapped atomic BEC's is given by
\begin{eqnarray}
i\hbar \frac{\partial \psi}{\partial
t}=-\frac{\hbar^2}{2m}\frac{\partial^2}{\partial x^2}\psi+V_{\rm{ext}}\psi+g|\psi|^2\psi-\mu\psi.
\end{eqnarray}
Here $\psi$ corresponds to the macroscopic order parameter of the system,
 $V_{\rm{ext}}$ is the necessary external confining potential and $m$ is the atomic mass.  
The effective 1D interaction coefficient $g$ arises by assuming an axially
symmetric elongated
three-dimensional system and integrating out the transverse ground state wavefunctions.  This leads to 
$g=2\hbar^2a/(ml_\perp^2)$, where $l_\perp$ is the transverse harmonic oscillator length and $a$ is the {\it 
s}-wave scattering length.
The validity of a purely 1D description of a quasi-1D system has been considered
elsewhere \cite{Jackson,Muryshev1,Muryshev2,Huang,Parker}.  Also, $\mu=gn$ corresponds to the chemical 
potential of the system, with the condensate density denoted by $n$.
In the absence of an external potential (i.e. $V_{\rm{ext}}=0$), Eq.~(1)
becomes exactly integrable and supports
dark solitary wave solutions \cite{Zakharov}.  On a uniform background density $n$, a dark soliton with
speed $v$ and position $(x-vt)$ has the analytical form,
\begin{eqnarray}
\psi(x,t)=\sqrt{n}e^{-i(\mu/\hbar)t} \left( \beta
\tanh \left[ \beta \frac{\left(x-vt\right)}{\xi}\right]+i\left(\frac{v}{c}\right)\right).
\end{eqnarray}
Here  $\beta=\sqrt{1-\left(\frac{v}{c}\right)^2}$, and the soliton speed $v$
is intimately dependent on both the soliton depth $n_d$ (with
respect to the background density), and the total phase slip $S$ across the centre.
It is given by
$v=\sqrt{n-n_d}=c\cos(S/2)$, the limiting value being set by 
 the Bogoliubov speed of sound 
$c=\sqrt{\mu/m}$.
A stationary soliton features a $\pi$ phase slip and a maximal depth 
$n_d=n$, while a soliton with 
speed $c$ is effectively indistinguishable from the background fluid.
The healing length $\xi=\hbar/\sqrt{\mu m}$ characterises the width of the soliton.

In the case of nonlinear optics, an equation analogous to the homogeneous version of Eq.~(1) arises
for nonlinear
media with $\chi^{(3)}$ susceptibilities, for which $\psi$ would correspond to the electric field amplitude.
Imperfections in $\chi^{(3)}$ crystals are unavoidable, and lead, in general, 
to perturbations of the 
susceptibility. This essentially introduces modifications to Eq.~(1), which arise, for example,
due to transiting, saturable or competing nonlinearities \cite{Kivshar}.
Any deviation from the homogeneous form of Eq.~(1) breaks its integrability and induces instability
in the soliton. As a result, the soliton emits radiation and
can subsequently decay. The rate of energy loss of the soliton
due to modified nonlinearities has been shown
to be  proportional to the local
acceleration squared \cite{Pelinovsky}. 
In the case of atomic BEC's,
the integrability of Eq.~(1) is broken due to the external confining potential
$V_{\rm{ext}} \neq 0$ within which experiments are performed.
In addition to the effects of transverse degrees of freedom
 \cite{Muryshev2,Huang,Parker,Feder,Brand,Komineas},
the integrability of the system is broken by the applied
longitudinal potential which induces the
soliton to radiate sound \cite{Huang,Parker,Busch,Fedichev}.

If the change in the background potential is weak, then,
to first order, the soliton dynamics are similar to those of
an effective
particle of constant (negative) mass \cite{Moura,Morgan,Reinhardt}.
In general, however, the sound emission leads to a change in the soliton
effective mass, and hence a change in its subsequent trajectory
\cite{Huang,Parker,Busch,Fedichev,Brazhnyi}.
To monitor this change, one can apply a perturbative approach \cite{Kivshar3,Konotop2}, which will
be valid as long as the potential does not vary significantly over the size
of the soliton.
However, if the density changes over a prolonged region, leading to continuous sound emission, then 
the total emitted energy can be a significant fraction of the initial soliton energy, and, consequently,
the {\it net} 
change in the soliton dynamics can be significant.
In previous work we have quantified the amount of sound emission and
the stability of dark solitons in quasi-1D BEC's, suggesting how this
can be both controlled and measured experimentally in realistic traps \cite{Parker}.
 In this paper we
 discuss in more detail the dynamics of solitons incident on
various inhomogeneities, with the aim to highlight the importance of the apparent {\it deformation}
that a soliton undergoes when incident on a potential gradient. 
 This deformation is a 
direct consequence of the sound energy trapped within the soliton region, as the soliton
propagates on the background density gradient.
The rate of emission of sound energy
by the soliton will be shown to be proportional to the square of the deformation parameter,
identified as the displacement of the observed soliton
centre of mass (i.e. including local sound energy within the soliton region) from the soliton (density) minimum. 
We will further 
show that the acceleration of a soliton through the fluid is directly proportional to the
soliton deformation parameter, thus
simultaneously confirming the validity of an acceleration squared sound emission 
law for various potentials.

This paper is structured as follows: Sec.~2 discusses the dynamics of dark
solitons incident on potential steps of variable length and height. This analysis
reveals the dissipative nature of the soliton motion and, in particular, highlights the significant proportion of the 
soliton energy that can be radiated as sound.
In this section, we further quantify the 
sound
emission as a function of the step geometry. The scattering of a soliton by a barrier of finite width 
contains a number of subtle features 
 which are discussed in Sec.~2.2.  The appearance of soliton deformation
becomes more apparent when considering the case of a constant background 
density
gradient, which can be provided by a linear ramp (Sec.~3). To discuss the
deformation in more detail, and to give results which are directly relevant
to atomic BEC experiments, Sec.~4 focuses on  a soliton
oscillating in a double gaussian trap.  This enables us to probe a large
range of background density gradients, and thus show that the
rate of sound emission can be parametrised in terms of the square of the induced soliton deformation.
 Finally, Sec.~5 contains some
concluding remarks.

\section{Dark Soliton incident on a potential step}

Consider initially a dark soliton on a homogeneous background 
incident on a potential step, as shown schematically in Fig. 1(a).
For an effectively infinitely long step of constant
(finite) height, this situation corresponds to soliton propagation between two homogeneous regions of 
different background density.  This is similar to bright soliton propagation at the interface between two 
nonlinear dielectric media of different refractive indices, arising in the context of nonlinear optics.  
This has been discussed within the effective (or equivalent) particle theory, which treats the bright 
soliton as a particle of fixed effective mass moving in an effective potential created by the interface 
between the two media \cite{Aceves}. Such an approach ignores radiative effects, 
which have however been included, to lowest order, in a perturbative treatment \cite{Kivshar4,Kivshar5}.

A soliton incident on a step of {\em finite} length can represent the presence of impurities, or
optically-engineered barriers.  Huang {\it et
al.} \cite{Huang2} have predicted that a dark soliton interacting with a
steplike potential emits radiation, and in the case where the step is downward, the
soliton is prone to fission into multiple solitons.  Frantzeskakis {\it et al} 
\cite{Frantzeskakis} recently considered the
interaction of a dark soliton with an impurity (without explicitly discussing sound generation),
whereas a subsequent paper \cite{Nistazakis}  addressed the possibility to trap solitons in time-dependent 
potentials.
Soliton-impurity interactions have also been discussed in \cite{Konotop}, where the 
expected
sound emission could not be observed due to computational constraints.

To study such systems in more detail, we consider a potential of the form,
\begin{equation}
V(x)=\left\{
\begin{array}{c}
0 \\
 V_0
\end{array}
\right. {\rm for}
\begin{array}{cr}
& x\leq 0, \hspace{0.1cm} x\geq L \\
& 0 < x < L
\end{array}
\end{equation}
where $V_0$ is the height and $L$ the length of the step.
In our subsequent numerical analysis, we employ dimensionless units, with distance
and velocity measured in terms of the healing length $\xi$ and the
speed of sound $c$ respectively.  In addition the asymptotic number
density $n_0$ is rescaled to unity.  The initial state is taken as
the product $\psi(x,0)=\sqrt{n(x)}\psi_{s}(x,0)$, with the soliton
centred on a locally flat background density.  Here $n(x)$ is the
time-independent background density obtained by imaginary time
propagation of the GPE, and $\psi_{s}(x,0)$ corresponds
to the homogeneous analytic soliton solution of Eq. (2).

\begin{figure}
\begin{center}
\hspace{0.125in}  
\includegraphics[width=8.2cm]{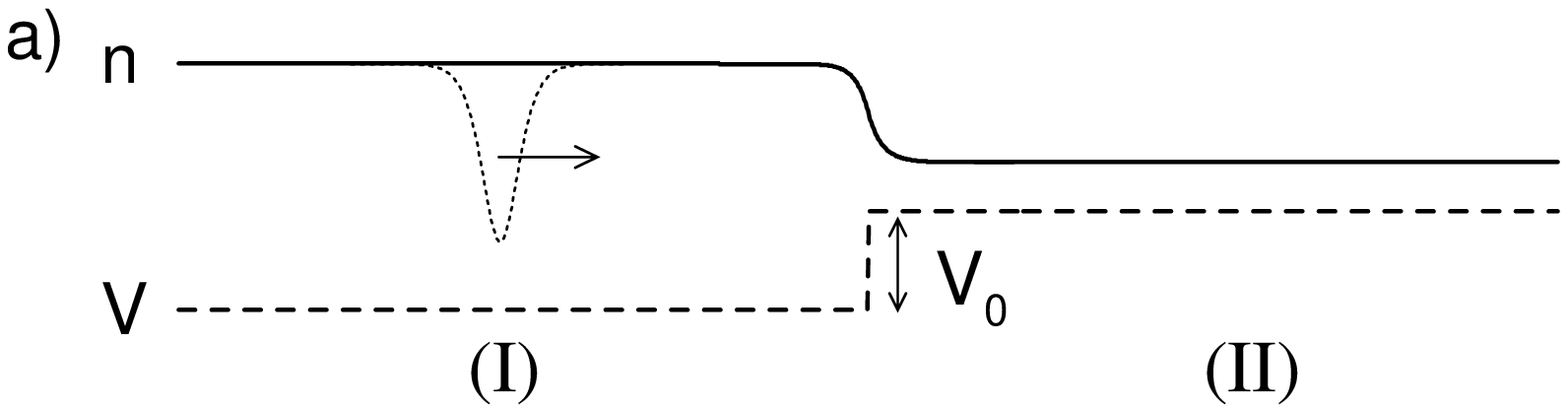}
\includegraphics[width=9.6cm]{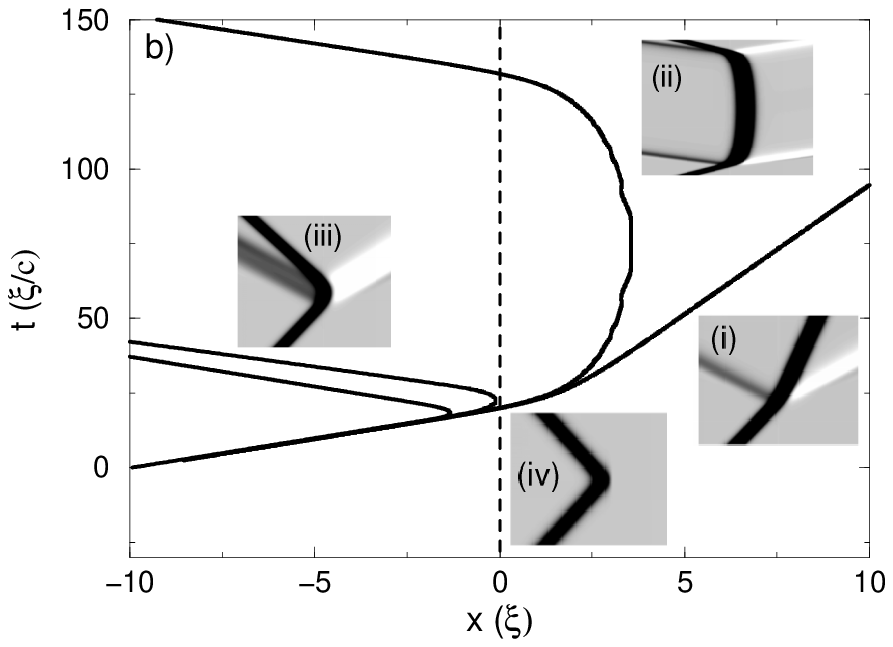}
\includegraphics[width=9.0cm]{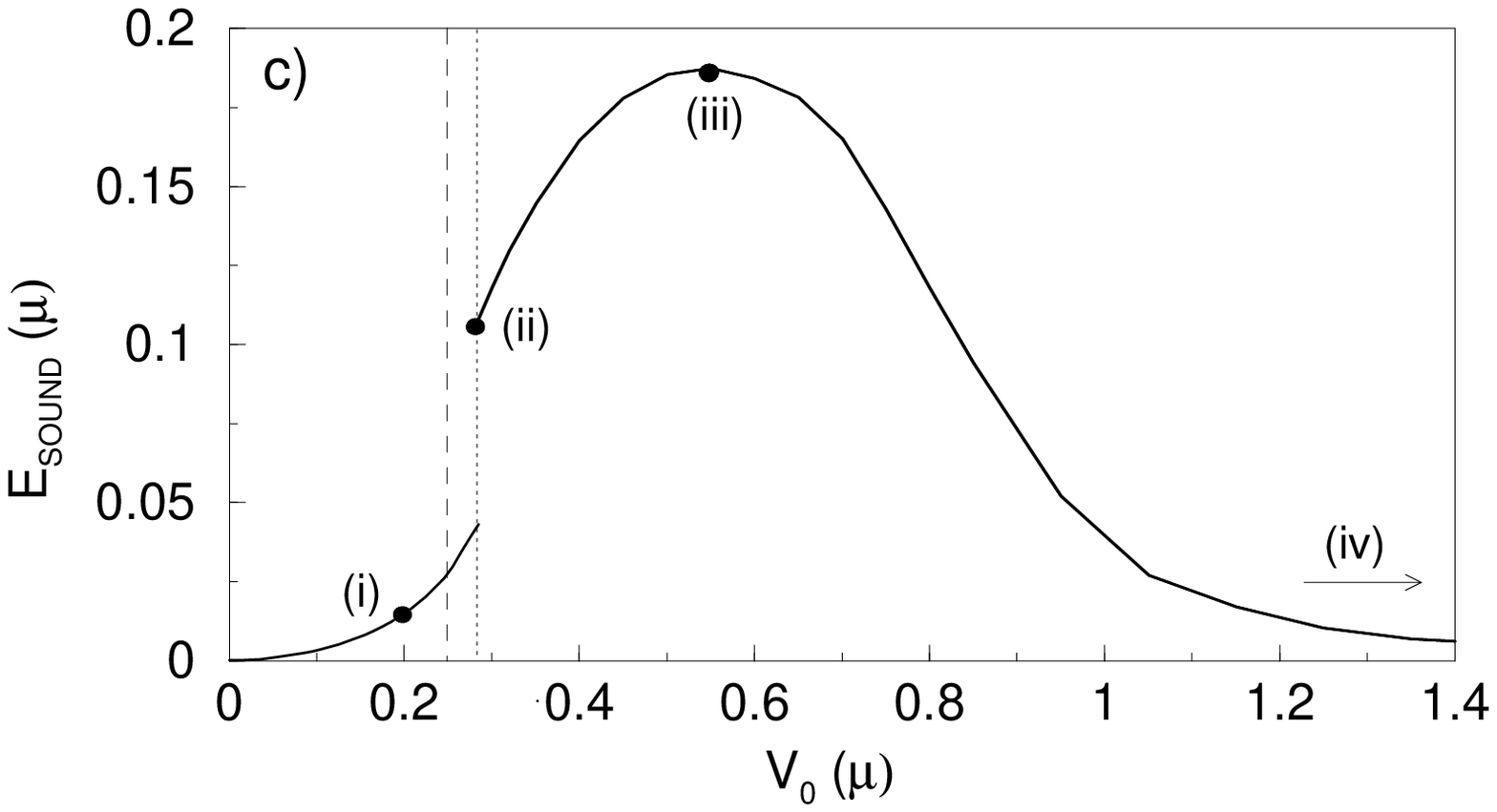}

\caption{
a) Dark soliton incident on an infinitely long step of height $V_0$ (dashed line).
The corresponding background fluid density is shown schematically by a solid line, while
the dotted line indicates the dark soliton propagating on the background fluid.
b) Path of a soliton, initially in region (I) with speed $0.5c$, 
incident on a
potential step of infinite length (region (II)).
We identify four regimes, in order of increasing step height: (i) transmission over the
step ($V_0=0.2 \mu$), (ii) reflective quasi-trapping at the boundary ($V_0=0.267 \mu$),
(iii) reflection with maximum sound emission ($V_0=V_0^{m}=0.55 \mu$), and iv) elastic collision
($V_0\gg\mu$), where $\mu=\hbar c/\xi$. 
Corresponding space-time carpet plots of renormalised density (i.e. by subtracting the time-independent background) for each case 
highlight
excitations on top of the background density,
with counter-propagating sound pulses clearly visible for cases (i)-(iii).
c) Total emitted sound energy as a function of step height $V_0$, 
with the cases (i)-(iv) from Fig.~1(b) 
highlighted. The point of reflection (represented by the dotted line) occurs for
larger step heights than predicted for an effective constant mass particle (dashed line), as a result
of sound emission.  The discontinuity arises from the fact that, during transmission a single burst of sound is 
emitted, 
while for reflection, there are two bursts, when the soliton passes both on and off the step.}
\end{center}
 \end{figure}

\subsection{Infinitely long step}

Consider initially
the case of a very long $(L \rightarrow \infty)$ positive $(V_{0} >0)$ step, as illustrated in Fig.~1(a).
This can be thought of as a transition between two different homogeneous regions.
To understand the ensuing soliton dynamics it is important to note that
the time-independent background density consists of the usual Thomas-Fermi 
solution, $n(x)=n_0(x)-V(x)$, 
plus a tanh-like smoothing of the boundary region resulting from fluid healing.
A positive step therefore induces a lower background density in the second region (II).
The different regimes of the dynamics of a dark soliton incident on such a boundary are
summarized in Fig.~1. In particular, Fig.~1(b) shows characteristic paths of a dark soliton with fixed initial 
speed
incident on steps of different heights $V_{0}$ (cases (i)-(iv), in order of increasing $V_{0}$).
Although this figure is based on a particular {\it initial} soliton speed, the qualitative
behaviour analysed here is characteristic of the general crossover between the two regions.

For low steps heights (Fig.~1b(i)) the soliton, initially in region (I), is able to pass over the
boundary,  and emits a burst of counter-propagating sound
pulses.  This sound serves to remove excess phase from the soliton as it is forced to 
adapt to the new ambient density. The radiation travels at the
speed of sound and
disperses over time. 
Once the soliton has fully traversed the barrier and entered the second homogeneous region, it propagates with reduced (but 
constant) speed, due to 
the lower ambient density and hence reduced speed of sound.
The emission of sound tends
to slightly increase the speed of the soliton, but this is only a minor effect compared to the
decrease in the soliton speed due to the lower background density.
Increasing the step height results in the soliton becoming ``refracted'' to a greater extent.
This ultimately leads to reflection, when the kinetic energy becomes
insufficient to ascend the potential (Fig.~1b(ii)),
with the amount of `penetration' into the second region decreasing with
increasing step height (Fig.~1b(iii)).
Assuming, as a first approximation, that both energy and depth of the soliton
remain conserved, the
threshold for transmission would be $V_0=v^2$, i.e., when the soliton
with depth  ($1-v^2$)
can {\it just} transmit into the far region with density ($1-V_0$).  However, the
process of sound emission tends
to reduce the soliton
depth and increase its speed. This leads to the appearance of a regime 
in which the soliton can be transmitted
even though the corresponding constant mass particle would be reflected (see next section).  
In this case sound is still radiated by the soliton
as it ascends and descends the bump.
Finally, in the limit of collision with a hard wall ($V_0 \rightarrow \infty$, i.e. soliton incident
on a region of zero density), the soliton reflects
elastically without breaking up or decaying, irrespective of the incident speed (Fig.~1b(iv)).
Note that the soliton reverses direction
before its centre
of mass actually reaches the boundary.
This is analogous to the
head-on collision of two solitons whereby momentum can be
exchanged without the centres having crossed, and
leads to the idea that the soliton undergoes a virtual collision
with its corresponding mirror image \cite{Huang2}. The fact that the macroscopic
wavefunction tends to zero at the wall implies a singularity in
the macroscopic phase, allowing the soliton to reset its phase
such that incoming and outgoing speeds are the same, which corresponds to elastic
reflection.

The analogy to refraction of light at an optical interface leads to
the question of whether there is a critical step height where the 
soliton becomes permanently trapped at the boundary.  We find a  
critical regime where the soliton becomes `quasi-trapped' at the
boundary (Fig.~1b(ii)), but that, within our resolution, the ultimate fate of the soliton is to either
transmit or reflect \cite{Frantzeskakis}. However, the   
soliton can indeed become {\it almost} stationary at the boundary for a   
relatively long time, during which it makes a significant passage
into the classically forbidden region. For example, for a soliton with initial speed $0.5c$ and a 
carefully selected step height 
$V_0=0.2666\mu$, we have observed quasi-trapping for times of $\sim1500(\xi/c)$ and the density minimum 
extending $\sim7\xi$ into the far region, before returning to the first region.  Two separate bursts of 
sound radiation are observed as the soliton passes on
and off the step. A closer inspection also reveals small-scale sound emission in the intervening time.

In the regime $V_0<0$, i.e., when the soliton travels to a higher density
region, we find similar results.  Here the soliton is always transmitted, sound is emitted, 
and the soliton speed increases due to the increased speed of
sound.  We fail to observe the
disintegration of the incident soliton into multiple solitons
predicted by Huang {\it et al} \cite{Huang2}.  In fact, we find that the soliton structure is
unbreakable by the potential step, and conclude that the main decay 
channel is
through sound emission.

The soliton energy can be calculated by integrating the energy
functional,
\begin{eqnarray}
\varepsilon(\psi)=\frac{\hbar^2}{2m}\left|\nabla\psi\right|^2+V_{\rm
ext}\left|\psi\right|^2+\frac{g}{2}\left|\psi\right|^4,
\end{eqnarray}
across the soliton region (say $x_s\pm5\xi)$ \cite{Carr1} and subtracting the corresponding contribution of
the background fluid.  This procedure cannot discriminate between soliton and sound energy
present in the interval, and this will be shown to be intimately linked to the
apparent soliton deformation.  The total emitted sound energy is shown in Fig. 1(c) as a function of step height, 
with the four characteristic cases of  
Fig.~1(b) highlighted.  
Its form is a result of the interplay between the magnitude of the density 
gradient experienced by the soliton and the reduced density in the second region, which tends to suppress sound transmission.  
For low step heights, the soliton 
has sufficient energy to transmit into region (II), emitting a single burst of sound energy as it passes onto the step.
Subseqeuntly the soliton remains in region (II), and the energy loss is determined by comparing the energy 
within the `soliton region' in regions (II) and the initial soliton energy.
As the step height increases, so do the depth and gradient of the background density 
perturbation,
leading to an initial increase in sound emission. At $V_{0} \approx 0.28 \mu$ (dotted line in Fig.~1(c)), the soliton no longer has
enough energy to be transmitted,
and thus eventually returns to region (I), after spending a finite time on, or near, the interface between the two 
regions. In this case, the final soliton energy is measured once the soliton has fully returned to region (I).
Hence, in this regime of quasi-trapping/reflection, the soliton emits two bursts of sound: one as it ascends the boundary, and another 
as is passes back off the boundary.  
This 
leads to a sudden 
increase (roughly doubling) of the 
emitted sound energy, as evident in Fig.~1(c).
The total sound emission initially increases further beyond this point, until the 
density gradient {\it experienced by the soliton} saturates.  
However the depth of the density inhomogeneity continues to increase, thus suppressing sound propagation into the second region and 
leading to a net reduction in the sound emission.
It is important to note that the total energy released  
can be considerable: for the case (iii) in Fig.~1, approximately $22\%$ of the soliton energy is converted into sound.

\subsection{Step of finite length}

\begin{figure}
\begin{center}
\includegraphics[width=8cm]{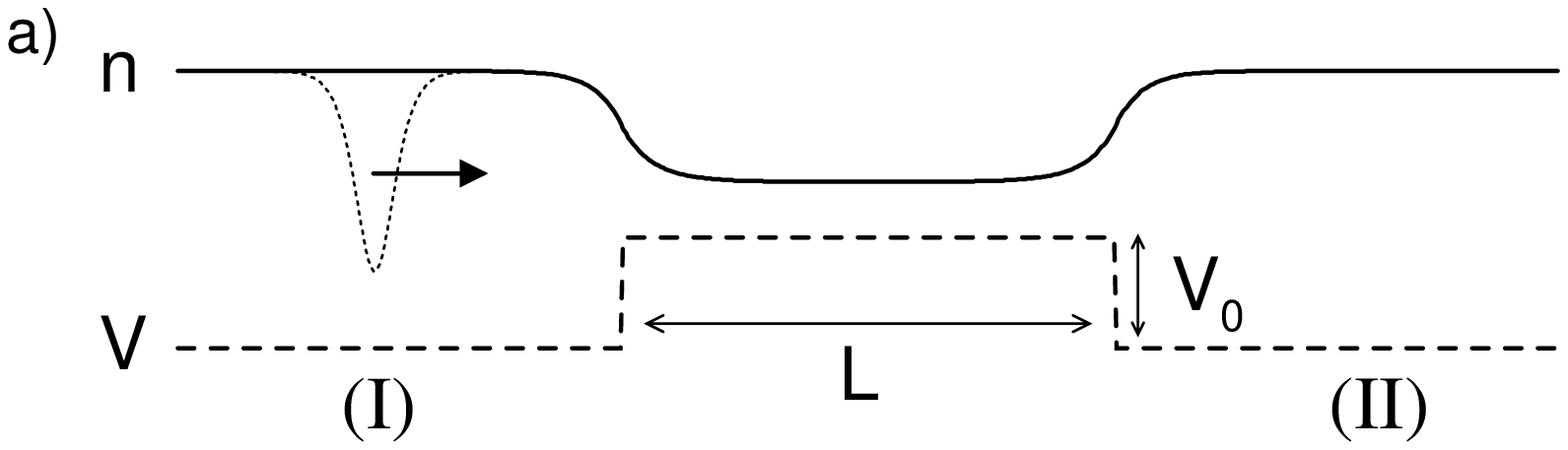}
\includegraphics[width=12cm]{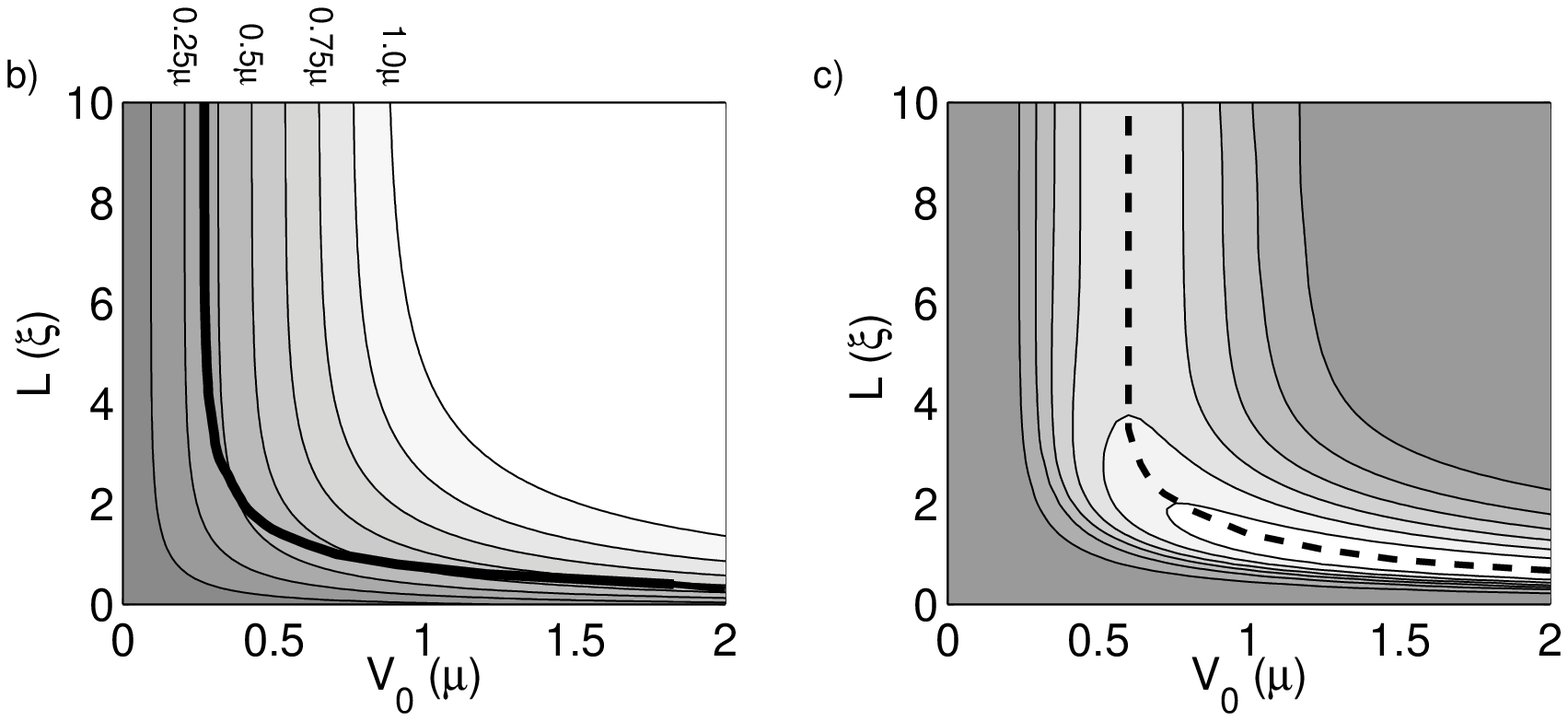}
\includegraphics[width=12cm]{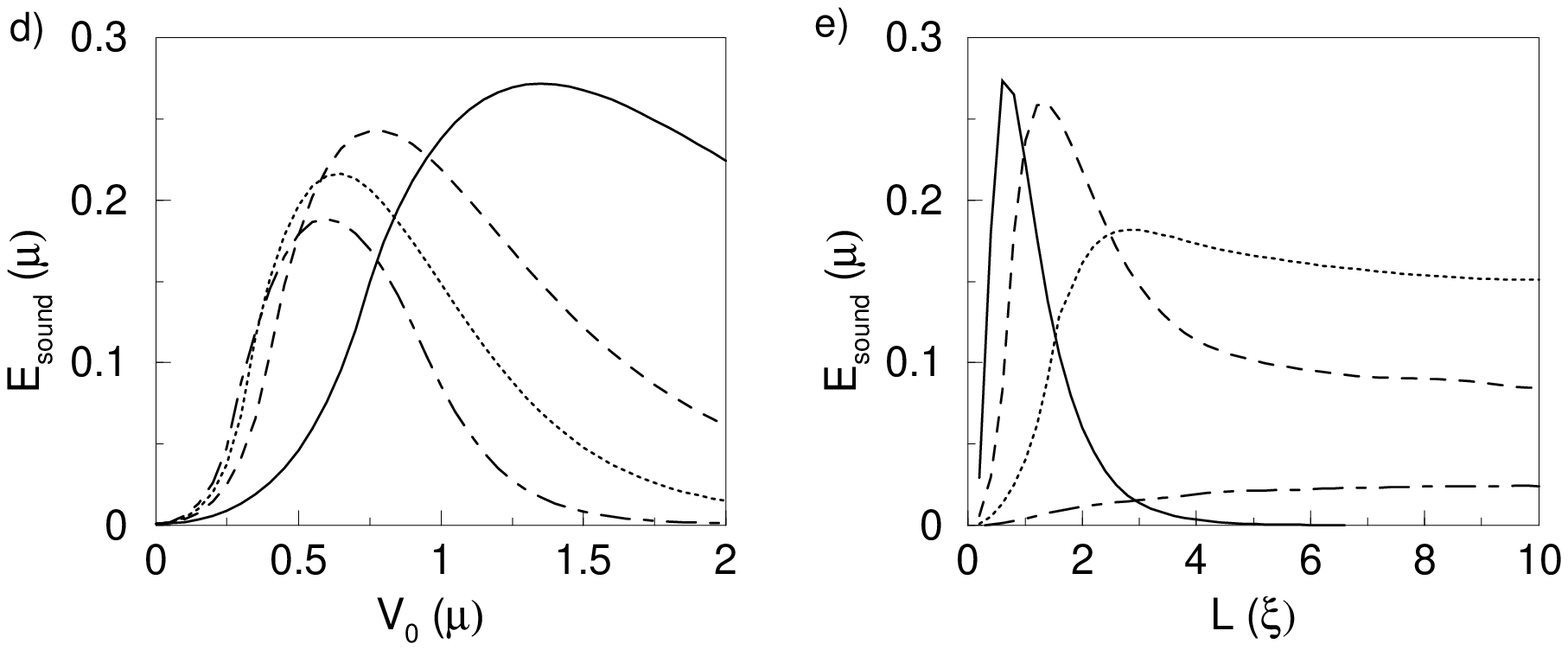}
\caption{ a) Soliton incident on a finite potential step (dashed line) of length $L$ and height $V_0$.
The step causes a reduction in the background fluid density
(solid line), through which the soliton (dotted line) propagates.
b) Effective potential imposed by the step as a function of length and height, with 
each contour representing an increment of $0.125 \mu$.  An effective potential of unity implies a pinning of the density to zero.  
Although 
higher effective potentials exist, we do not show them here as, throughout this limit, the sound emission is essentially prohibited.  Note 
that for 
effective potentials slightly greater than unity, the density is weakly pinned to zero and sound of positive amplitude can just propagate 
through.
The solid line
indicates the transition between reflection/transmission for a soliton with initial speed $v=0.5c$ (the
non-dissipative prediction would be the $V_{\rm eff}=0.25 \mu$ contour). c) 
Total sound emitted from a dark soliton with
inital speed $v=0.5c$ and energy $E_s=0.866\mu$ incident on a step of height $V_0$ and length $L$.
The emitted energy ranges from zero (darkest region) to maximum value in equal contour steps.
Maximum sound emission occurs along the dashed line $V_0=V_0^{m}(L)$. 
d) Cross-sections of c) for constant
step lengths of $L=\xi$ (solid line), $2 \xi$ (dashed line), $5 \xi$ (dotted line) and $10 \xi$ (dot-dashed line). 
e)  Cross-sections of c) 
for 
constant step heights of $V_0=0.25 \mu$ (dot-dashed line), $0.5 \mu$ (dotted line), $\mu$ (dashed
line)
and $2 \mu$ (solid line).} 
\end{center}
\end{figure}

To perform a more detailed study of the dissipative soliton dynamics,
we now consider a dark soliton incident on a step of finite length $L$,
as illustrated in Fig.~2(a).  The
important parameter determining whether a soliton is transmitted at a barrier is
the `effective' density seen by the moving soliton due to fluid healing in the vicinity of the step
(solid line in Fig.~2(a)).
For sufficiently narrow steps, fluid healing prohibits the density from reaching the Thomas-Fermi value, $n(x)=n_0(x)-V(x)$.
The minimum density $n_{\rm min}$ to which the fluid actually heals can be used to define an effective potential of height $V_{\rm 
eff}=(1-n_{\rm min})$.
 Fig.~2(b) maps the effective potential for steps of variable
length and height. This suggests that for small enough step widths, the 
soliton 
may
see an effective step height much lower than the actual potential, thus 
enabling the soliton to tunnel through barriers 
for which a classical particle would undergo elastic reflection.
Assuming no dissipation, then, to leading order, the requirement for transmission would be that
$V_{\rm eff} < v^{2}$. Thus, for a soliton
with incident speed $v=0.5c$, the $V_{\rm eff}=0.25 \mu$
contour of Fig.~2(b) represents, to first order, the critical region between transmission and reflection.
The {\it actual} crossover between
reflection and transmission is shown by the
solid line in Fig.~2(b), which clearly shows the deviation resulting
from the inhomogeneity-induced emission of sound from the soliton. 

Fig.~2(c) shows the corresponding total sound emitted from the soliton,
the magnitude of which is intimately related to the above-mentioned effective
potential.
Maximum sound emission occurs in the region of reflective quasi-trapping
 where the soliton is subject to its greatest acceleration/deformation.
The line of maximum sound emission, defined by $V_0=V_0^m(L)$, is highlighted by the
dashed line in Fig.~2(c).  Note that $V_0^m(L)$ decreases with increasing length, and essentially saturates to 
the value of $V_0^m(L)\approx 0.6\mu$
for $L\geq
5\xi$.
The maximum sound energy released in Fig.~2(c) is approximately $0.27\mu$, which corresponds to $32$$\%$ of the 
initial soliton 
energy.  
The sound emission tails off in the opposite limiting cases of
$V_0,LEq.(4)_0\rightarrow0$ and $V_0,L_0\rightarrow\infty$.
To visualize the behaviour more clearly, Fig.~2(d)-(e) plot, respectively, the total
emitted sound energy (i.e. both forward and backward sound pulses) as a function of step height $V_0$
(for various fixed step widths $L$) and as a function of $L$ (for various fixed $V_0$).

Fig.~2(d) shows that, for fixed $L$, the total emitted sound increases abruptly with 
increasing step height, reaches a maximum at the point of reflective quasi-trapping,
and decreases more slowly as $V_0 \rightarrow \infty$, similar to the behaviour encountered in 
the last section. 
Note that the variation of the emitted energy with step height is smooth and does 
not feature the discontinuity apparent in Fig.~1(c) 
for an infinitely long step.  This can be understood as follows: for a soliton transmitting 
over an 
{\it infinitely} long step, a single burst of 
energy is emitted before the soliton 
reaches its steady state on the step, whereas a soliton which is ultimately reflected interacts {\it twice} with the
boundary, thus emitting two sound pulses.  
In the case of a {\it finite} length step, this distinction does not arise.
For transmission, the soliton must ascend and descend 
the step before reaching its steady state 
on the far side of the step (where its final energy is to be computed). 
This clearly involves two bursts of energy being radiated.  
However, the case of reflection also features two bursts, occuring when the soliton
ascends and descends the first boundary. 
Since the background fluid is symmetric around the centre of the step
(as regions (I) and (II) have the same potentials), the net amount of sound emission
in the two limiting cases of soliton marginally transmitting, or just reflecting,
will be the same, and hence the discontinuity will be absent here.

To understand in more detail the features of Fig.~2(d), it is instructive to consider the effective potential
experienced by the soliton in each of the three limiting cases, namely
(i) transmission ($V_{0} \ll V_{0}^{m}(L)$),
(ii) reflection with maximum energy emission (occuring at $V=V_{0}^{m}(L)$), and (iii) reflection with sound emission 
($V_{0} \gg V_{0}^{m}(L)$), but still far from the  $V_0 \rightarrow \infty$ limit.
These are shown in figures ((a)-(c)), respectively.  Different lines in each of these figures correspond to 
different
step widths ($L=\xi$, $2\xi$, $5\xi$ and $10 \xi$).  (Note that the step heights in Fig.~3(b) differ from line 
to line.)
For a constant step width $L$, the density depression becomes more pronounced with increasing step height
(this can be visualized, e.g. by looking at the change of the solid line corresponding to $L=\xi$ from (a)-(c)).
 Fig.~3(a) shows the case of 
soliton transmission over the 
entire step ($V_0=0.25\mu$). Note that the particular soliton under consideration, with initial speed $0.5c$ and depth $0.75n_0$, 
can probe, to first order, 
densities down to $0.75n_0$ before being forced to change direction.  Such a soliton can therefore just transmit in this case.  In 
this limit 
of transmission,
maximum sound emission occurs when the soliton traverses the steepest background density gradient
for the longest time. Clearly this occurs for maximum step width $L=10\xi$ (dot-dashed line).
Fig.~3(c) shows the other extreme case, when the soliton is completely reflected, as a result
of experiencing a very high potential ($V_0=2\mu$). 
In the case of a sufficiently long step, e.g. $L=10\xi$ (dot-dashed line) in Fig.~3(c), the fluid heals fully to the 
Thomas-Fermi limit, thus reaching zero 
density.  The soliton effectively sees a hard wall and becomes reflected without any sound emission (see corresponding dot-dashed 
line in Fig.~2(d) at $V_0=2\mu$).
However, as the step width is decreased, the reflected soliton can emit increasing amounts
of sound energy.  This is because the finite density at the step now enables {\it sound propagation} into the second 
region, despite the fact that the {\it soliton} is 
reflected.  
Note that, despite appearing to do so, the dotted line of Fig.~3(c) does {\it not} reach zero density
(see Fig.~3(e) showing enlarged version of Fig.~3(c) near $n=0$).
This explains why there is some (albeit limited) sound emission for $L=5\xi$ (dotted line) at $V_0=2\mu$ in Fig.~2(d).

\begin{figure}
\begin{center}
\includegraphics[width=12cm]{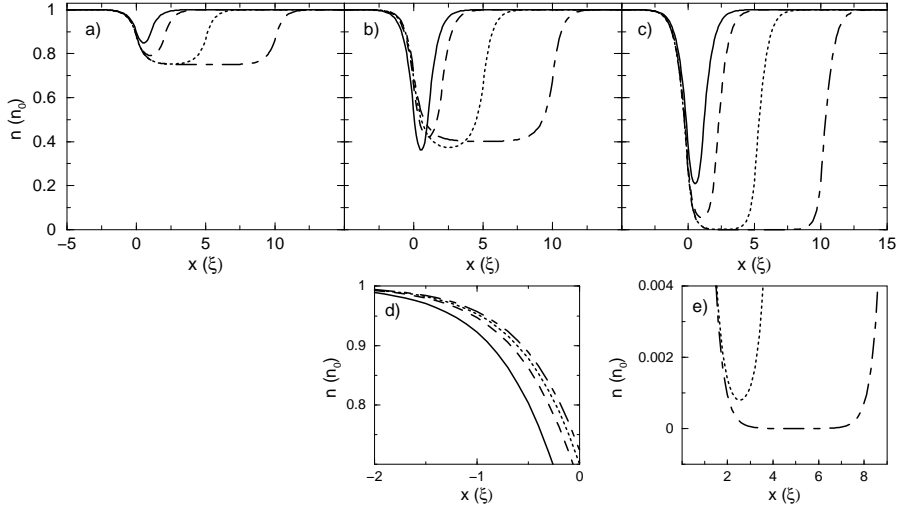}
\caption{Background density profiles due to fluid healing in the region of the step.
Each sub-figure plots the cases of 
$L=\xi$ (solid line), $2 \xi$ (dashed line), $5 \xi$ (dotted line) and $10 \xi$ (dot-dashed line)
as in Fig.~2(d) for different potential heights $V_{0}$:
(a) Transmitted soliton at $V_{0} = 0.25 \mu$, 
(b) Reflected soliton at $V_{0}=V_{0}^{m}(L)$ corresponding to maximum sound emission  for each $L$, 
i.e. at $V_0^{m}(\xi)=1.36\mu$ (solid line), $V_0^{m}(2\xi)=0.77\mu$ (dashed line), 
$V_0^{m}(5\xi)=0.64\mu$ (dotted line) and $V_0^{m}(10\xi)=0.60\mu$ (dot-dashed line).  
(c) Reflected soliton at $V_{0} = 2 \mu$. 
(d) Enlarged image of
the background density inhomogeneity experienced by the soliton in (b), in the region $-2<x<0$
for $L=\xi$, $2\xi$, $5\xi$, and $10\xi$. In this region, the density 
gradient is clearly
maximized for the narrowest step ($L=\xi$, solid line),
(e) Enlarged image of (c) showing the effective minimum density in the cases $L=5 \xi$ (dotted) and $L=10 \xi$ (dashed). 
This shows that for a width $L=5 \xi$ (dotted line), the density does not 
quite heal to zero, allowing a small amount of sound transmission through the barrier.
Note that such density profiles  can also be obtained by matching analytically known solutions of 
the Gross-Pitaevskii equation in finite and semi-infinite intervals (see, e.g. \cite{Carr2,Carr3}).

}
\end{center}
\end{figure}

Although this accounts for the qualitative features of the lines in Fig.~2(d), it still does not 
explain why the 
{\it maximum} emitted
energy for narrow steps (e.g. solid line in Fig.~2(d) at $V_0=V_0^{m}(L=\xi)\approx1.4\mu$) is larger than the 
{\it maximum} emitted energy for longer steps 
(e.g. dashed line of same figure). To explain this, we must examine the background density profile for each of the lengths considered 
(i.e. $L=\xi$, $2\xi$, $5\xi$ and $10\xi$) at the step height 
which corresponds to maximum sound emission.
As evident from Figs.~2(c)-(d), this critical barrier height $V_{0}=V_{0}^{m}$ depends also on the step width
(i.e. $V_{0}=V_{0}^{m}(L)$).
Therefore, to interpret the increase in the maximum emitted energy for narrower steps, we must compare the
effective background densities for different widths $L$ at their corresponding heights $V_{0}=V_{0}^{m}(L)$.
This is illustrated in Fig.~3(b). To first approximation, in all these 
cases (i.e. all $L$), the
soliton probes the step region down to a density of approximately ($1-v^2$) 
(corresponding to $0.75n_0$ for the soliton 
under consideration), before 
being reflected. Consequently, the soliton is approximately 
limited to the $x<0$ region.
Therefore, the key parameter controlling the amount of sound radiation is the {\it gradient} of the background density probed by 
the 
soliton (which clearly depends on the step width).
To see this gradient clearer, Fig.~3(d) plots an enlarged version of the effective potentials of Fig.~3(b)
in the region $-2 \xi < x < 0$. 
Hence, at the point of maximum emission for each $L$, it is actually the {\it narrower} step (solid line) which has the 
{\it steepest} gradient. As a result, the {\it maximum} sound emission occurs for narrower steps. We have hence accounted for the entire 
behaviour of the emitted sound energy as a function of step height.

From the above analysis, one can also understand the features of Fig.~2(e)
showing the dependence of emitted energy on step width. 
For small $V_0$ (e.g. $V_0=0.25\mu$, dot-dashed line) the soliton is 
transmitted, releasing maximum sound for longest 
steps, with the amount saturating for 
$L\rightarrow \infty$.
Increasing $V_0$ (e.g. $V_0=0.5\mu$, dotted line) leads to larger density 
gradients and hence more sound emission as 
$L\rightarrow \infty$.  The peak emitted energy 
at $L\approx 2.7\xi$ is due to a maximum time spent on the step in the 
regime of reflective quasi-trapping.  Increasing $V_0$ 
further leads to the gradual 
suppression of sound transmission and hence a reduced total sound emission in the limit $L\rightarrow \infty$.  
Note that as $V_0$ is increased, the phenomenon 
of quasi-trapping appears at smaller $L$, leading to a shift of the peak emission to the left.  The maximum 
sound emission occurs for large $V_0$ (and small enough $L$) since then the density gradient experienced by the soliton is maximised.  
This 
trend appears to be maintained with indefinite increase in $V_0$, although 
it requires smaller step widths, 
which will ultimately 
become 
unphysical. 
When $V_0>\mu$, the soliton undergoes elastic reflection with no sound emission, provided the step width $L$ is 
sufficiently long to allow for complete fluid healing to the imposed potential $V_0$.  This explains the tailing 
off of the sound energy for $V_0=2\mu$ (solid line) to zero as $L>5\xi$.

Our analysis so far has been based on a soliton of fixed initial speed, and this has enabled us to map
the entire dependence  of emitted sound energy on $L$ and $V_0$ for a particular soliton speed.
We shall now briefly comment on the dependence of the emitted energy on initial soliton speed.  
Fig.~4 shows the energy radiated by a soliton incident on a finite potential step 
of fixed length $L=5\xi$ and variable height $V_0$,
for various initial soliton speeds, with the energies scaled in terms of the initial soliton energies. In
the regime of transmission over the step
(i.e. leftmost part of the figure), the 
slowest soliton emits the most sound energy.  This can be attributed to the fact that, although all three solitons 
experience the 
same density gradient, the slowest soliton (solid line) spends the longest time at the step, thereby leading to the 
greatest time-integrated 
sound emission.  
However, the slowest soliton features the deepest density profile.  Its passage into the step region is 
therefore most restricted, causing it to be the first to both reflect and reach the point of saturation of the sound emission
(i.e. $V_0=V_0^{m}(L)$).
Thus, increasing initial soliton speed leads to a general shift of the emission curves to higher potentials.
Faster solitons can pass a greater distance into the step region, probing higher density gradients, and so 
have higher peak sound 
emission.  
However, the energy of an unperturbed
 soliton on a homogeneous background is given by \cite{Kivshar},
\begin{eqnarray}
E_{\rm sol}=\frac{4}{3}\hbar c n\left[1-\left(\frac{v}{c}\right)^2\right]^{3/2}
\end{eqnarray}
and hence, faster solitons have lower energies.  This means that, in addition to radiating more energy, faster solitons 
radiate a higher fraction of their total energy.  
For example, a soliton with initial speed $v=0.7c$ can radiate as much as $65\%$ of its initial energy.
In all cases the sound emission eventually tends to zero as the step height $V_0$ becomes much greater than the chemical 
potential, although faster solitons still emit more energy in the limit of high barriers.

\begin{figure}
\begin{center}
\includegraphics[width=10cm]{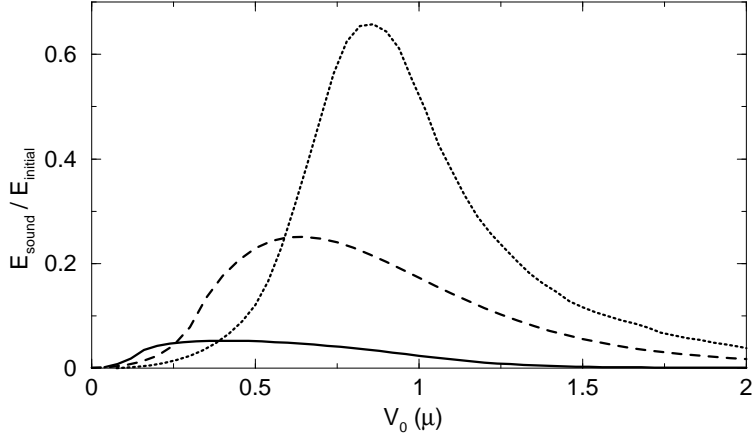}
\caption{Fraction of the initial soliton energy that is emitted for a soliton incident on a 
finite potential step of length $L=5\xi$ as a function of step 
height $V_0$ for various initial soliton speeds: $v=0.3c$ (solid line), 
$0.5c$ (dashed line - this corresponds to dashed line in Fig. 2(d)), and $0.7c$ (dotted line).
The initial soliton energies are $1.157\mu$, $0.866\mu$ and $0.486\mu$ respectively.
}
\end{center}
\end{figure}

We note that for effectively finite step sizes the soliton undergoes
considerable physical deformation as it interacts with the
barrier, due to the continuous sound emission. In particular the soliton profile becomes
asymmetrically distorted from its equilibrium state, which signifies a
deviation of the centre of mass from the density minimum.  
A detailed quantitative study of the deformation is limited by
the rapid change of the soliton profile when it is
incident on the barrier. To illustrate these effects more clearly,
we therefore turn our attention to linear ramps, for which the soliton,
once on the ramp, experiences a constant force.

\section{Soliton under uniform acceleration}

A linear potential provides a simple way of imparting a controlled and constant force to the
soliton.
As the soliton ascends the ramp, its depth stays (approximately) constant while the ambient density constantly decreases.
At
the point where the node of the soliton touches zero density, the soliton reverses direction, and begins to descend the 
ramp.  We observe
continuous sound emission from the soliton while it is on the ramp.  This causes small
deviations
from the corresponding classical trajectory, as shown in Fig. 5.

  \begin{figure}
\begin{center}
\includegraphics[width=8.0cm]{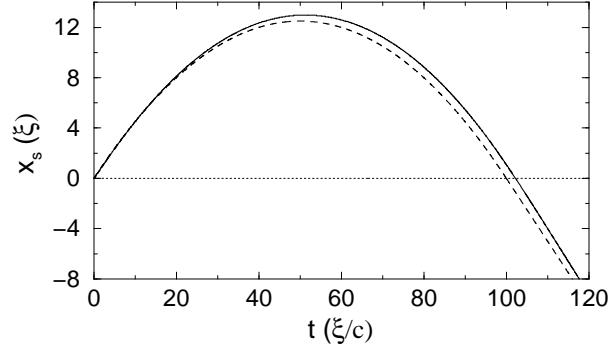}
\end{center}
\caption{ Path of a soliton (solid line), with initial speed $v=0.5c$, ascending and subsequently descending a
linear potential ramp
$V=2\times
10^{-2}x$ acting in the $x>0$ region. Due to the sound emission this
trajectory deviates from the constant mass effective particle prediction (dashed
line). } \end{figure}

\begin{figure}
\begin{center}
\includegraphics[width=10.5cm]{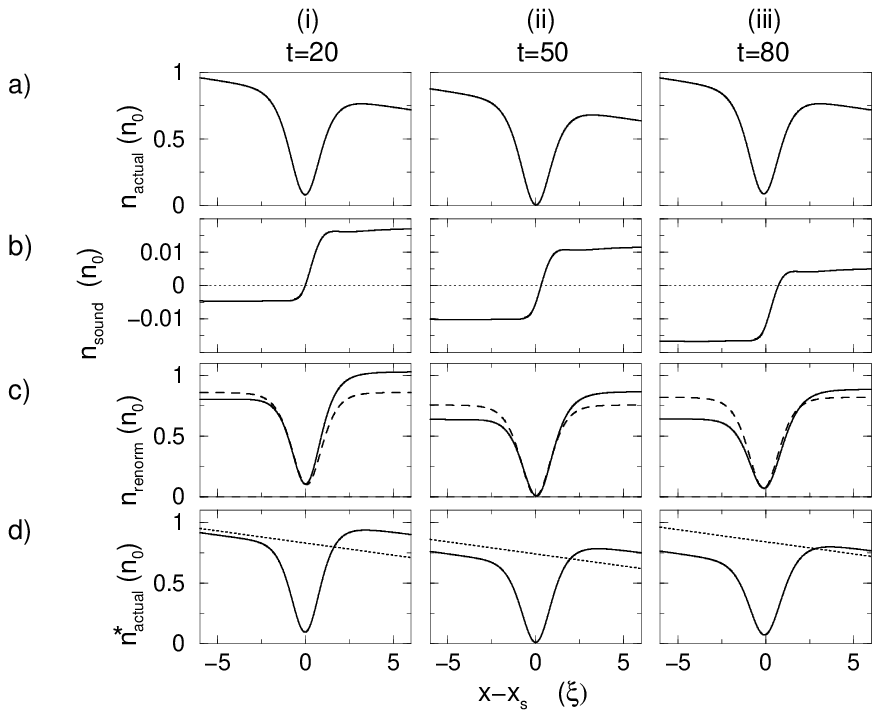}
\end{center}
\caption{a) Actual density profile of the soliton region, for the case of propagation on a linear ramp with the same 
parameters as Fig.~5, during (i) ascent 
($t=20\xi/c$), (ii) the turning
point ($t=50\xi/c$), and (iii) descent ($t=80\xi/c$).
b) Sound density within the soliton region, obtained as follows:  Firstly, we subtract the background density 
from the actual soliton profile, to obtain a `renormalised' soliton profile. Then, by subtracting from this 
the 
corresponding {\it unperturbed} soliton with the same speed and position (dashed line in (c)), we isolate the 
sound density.  The renormalised soliton profile, with the sound density (b) magnified by a factor of ten, is 
shown by the solid line in (c), and clearly illustrates the sound-induced deformation.  d) Same as (a) but, 
again, with the sound 
density
magnified by a factor of ten (solid line).  The dotted line indicates the 
corresponding time-independent background 
density.} \end{figure} 

Shapshots of the perturbed soliton and the corresponding sound density, for 
a soliton (i) ascending up the ramp, (ii) at the stationary point, and (iii) during descent are shown in Fig.~6.
The sound density (Fig.~6(b)) consists of a
higher density in front of the soliton and a lower density behind it, the fore and aft
pulses being spatially compressed and extended depending on the relative motion of the soliton
and sound through the fluid.  Note that, at the stationary point, where there is no relative motion between soliton
and sound field, the fore and aft density pulses have equal (but opposite) amplitudes.
The asymmetrically emitted sound within the soliton region leads to an apparent deformation of the soliton profile.  This
deformation can be visualised by comparing the renormalised soliton (i.e. soliton profile minus background density) to a 
corresponding unperturbed soliton of the same speed and position (Fig.~6(c)).  This deformation 
leads to a shift of the soliton centre of mass $\bar{x}$ from the 
density minimum $x_{s}$ \cite{comment}, where
the soliton centre of mass is defined as
\begin{eqnarray}
\bar{x}=\frac{\int_{\rm s} x \left(|\psi|^2-n\right) {\rm d}x}{\int_{\rm s}
\left(|\psi|^2-n\right) {\rm d}x}.
\end{eqnarray}
and the `soliton
region' ${\rm S}$ is taken to be ($x_s\pm5\xi)$. For a soliton on a ramp,
this yields a constant shift of the centre of mass from the soliton density minimum \cite{Frame}, 
indicating a link between soliton deformation
and acceleration.

The exception to this link arises when the whole fluid itself is accelerated.  The 
fluid 
motion then also accelerates the soliton, from the viewpoint of a stationary observer,
but the soliton is not accelerated with respect to the background fluid.  This could be achieved, for example, by displacing
the confining potential such that the whole system (background fluid plus soliton) move together.  Such motion does not induce a deformation
of the soliton profile and therefore does not lead to sound emission.  From now on, when referring to acceleration we therefore mean acceleration with 
respect to the background fluid.

\begin{figure}
\begin{center}
\includegraphics[width=6.2cm]{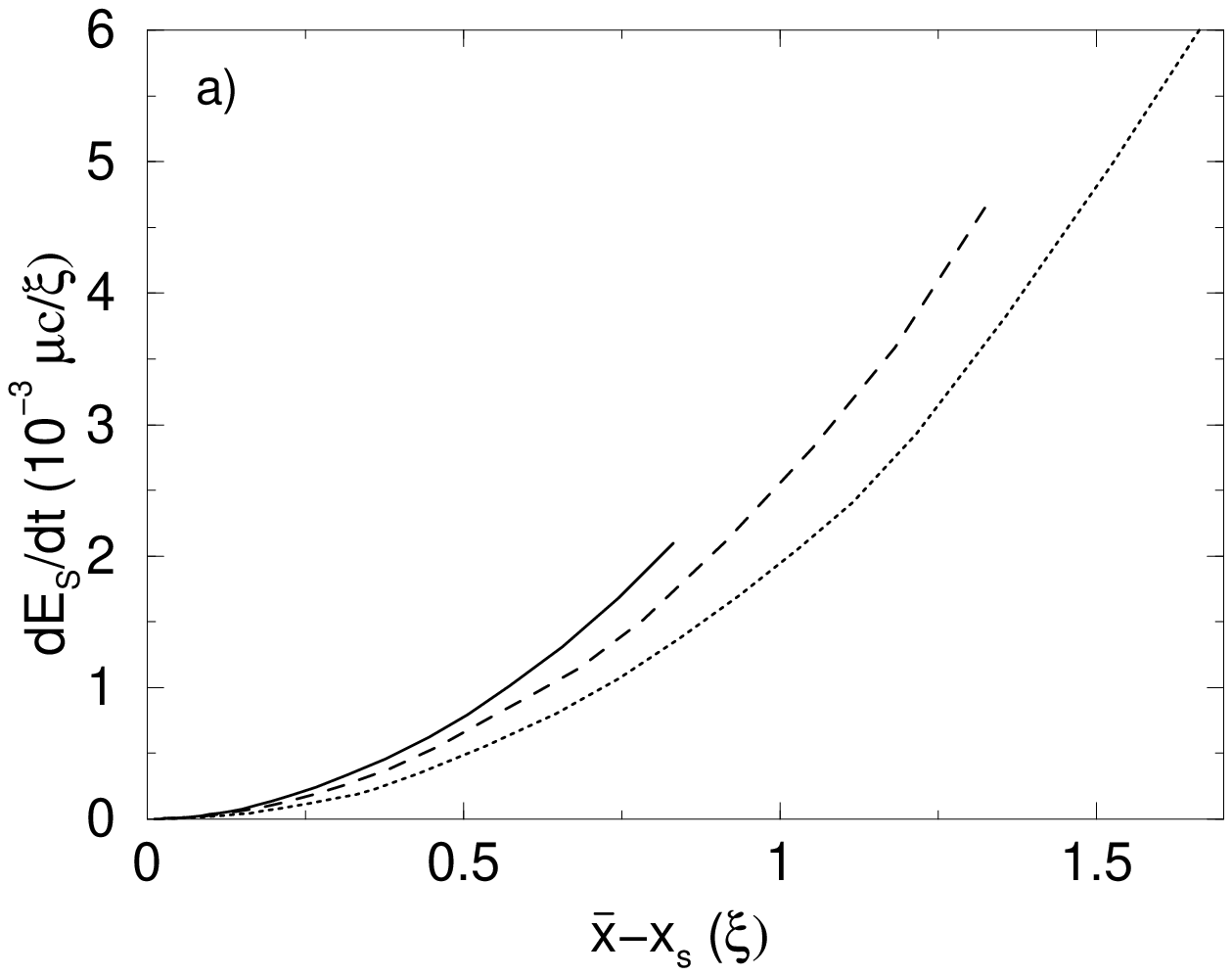}
\includegraphics[width=5.8cm]{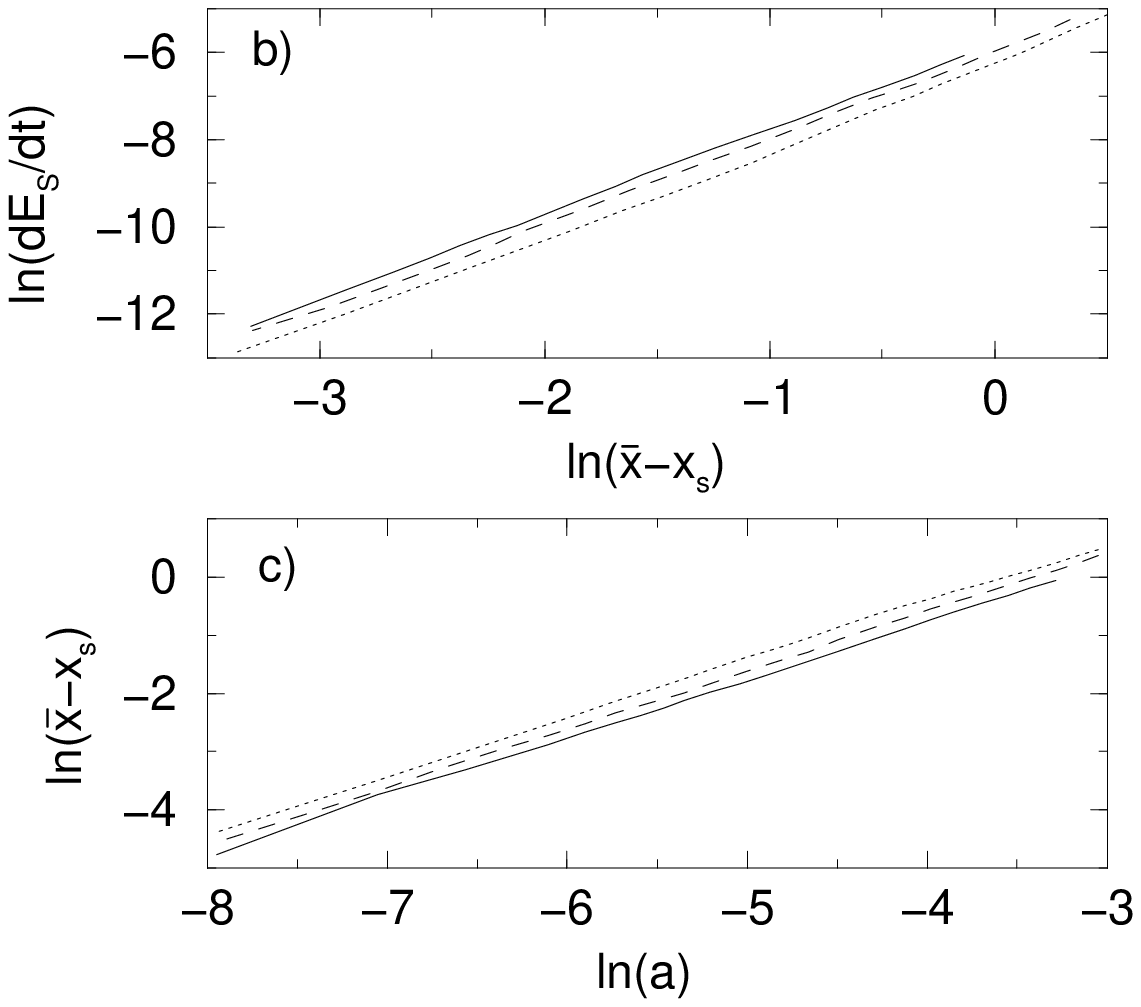}
\caption{ a) Power radiated from a soliton ascending, and subsequently 
descending,
a linear ramp, plotted versus the soliton deformation $(\bar{x} - x_{s})$.
Different curves correspond to different initial soliton speeds 
$v = $ $0.3c$ (solid line), 0.5c (dashed line) and 0.7c (dotted
line). Each curve has been constructed by probing over different
potential gradients, in the range $(10^{-3} - 10^{-1}) \mu/\xi$.
Note that solitons with larger initial speeds ascend higher up the ramp and the maximum
observable deformation increases with increasing initial soliton speed.
b) Same data plotted on log-log scale. c) Soliton deformation versus
acceleration (measured in units of $(c^2/\xi)$) on log-log scale, for the above mentioned soliton ascending a linear ramp.}
\end{center}
\end{figure}

The power radiated by the soliton while on
a ramp is
shown in Fig.~7(a), as a function of the deformation parameter $(\bar{x} - x_{s})$, for various initial soliton 
speeds.
This suggests a behaviour of the form
\begin{eqnarray}
\frac{{\rm d}E_{\rm s}}{{\rm d}t}=-\kappa(\bar{x}-x_{s})^{\alpha}
\end{eqnarray}
where the value of $\kappa$ is expected to depend, at least, on the 
initial soliton speed and the gradient of the local density. Fig.~7(b) plots the same
information on a logarithmic scale, thereby confirming the above 
behaviour. 
From this we determine, to very good approximation, the value $\alpha = 2$.
Fig.~7(c) shows, on logarithmic scale, the relation between
the deformation parameter and the soliton acceleration, for the case of a
soliton ascending a ramp. This points to a direct proportionality
between the two quantities, which is consistent with an acceleration-squared law
for emitted power \cite{Parker,Pelinovsky}.

\section{Soliton oscillating in a gaussian trap}

The relation between emitted power and soliton deformation can be put on firmer ground by
considering the case of an oscillating soliton, since such a case includes a wide range
of soliton speeds and background densities. The most lucid example is that of a symmetrical trap
formed by two gaussian-shaped potentials,
\begin{equation}
V=V_0\left[\exp\left\{-\frac{(x-x_0)^2}{w^2}\right\}+\exp\left\{ 
-\frac{(x+x_0)^2}{w^2}\right\}\right],
\end{equation}
within which the soliton is initially trapped, as illustrated in Fig.~8. Here $V_0$ is the amplitude of the gaussians, 
$x_0$ the distance of the centres from the origin, 
and $w$ the gaussian width.  To monitor the effect of sound emission
(without allowing for reabsorption of sound to complicate matters \cite{Parker}) we assume that the
height of the gaussian bump is low enough ($V_{0} \ll \mu$), such that the emitted sound escapes the system.
To avoid reabsorption of sound, one would need to  embed this double gaussian trap 
 in a
weaker (harmonic) outer trap. Then, as the soliton oscillates in
the  (inner) double gaussian trap, it emits counter-propagating sound waves which,
 for sufficiently low $V_{0}$,  `escape' to the outer trap
and do not re-interact with the soliton (within the timescales of interest). This results in
 a drastic change in the soliton amplitude and frequency.

In the last section we argued that the deformation of the soliton is 
intimately linked with the continuous sound emission, and gave evidence to suggest that the power emitted by the soliton
can be parametrised by the square of
the soliton deformation.   However, the coefficient $\kappa$ in Eq.~(7) 
will {\em not}, in general, be
a constant depending {\em simply} on {\em initial} soliton speed, due to the changing
background density gradient in the trap. We therefore anticipate $\kappa$ to 
depend on the position of the soliton in the trap 
(and hence become a function
of the local background speed and density)
in a non-trivial manner.

\begin{figure}
\begin{center}
\includegraphics[width=8cm]{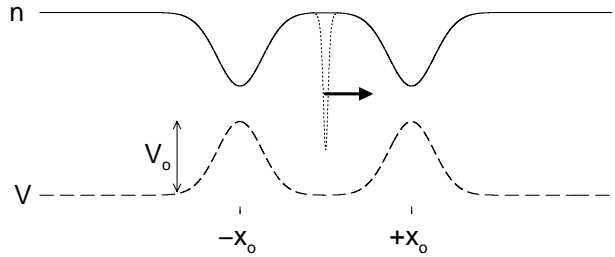}
\end{center}
\caption{A dark soliton (dotted line) is confined to oscillate within a trap formed by two identical gaussian bumps 
(dashed line) of height $V_0$ centred at $-x_o$ and $+x_0$.  The solid line represents the corresponding time-independent 
background density.} 
\end{figure}

Fig.~9 considers the long term dynamics of a soliton with initial speed $0.3c$ oscillating in a double gaussian trap.  The 
radiation from the 
oscillating soliton 
 leads to an `anti-damping' effect \cite{Busch}, whereby the soliton oscillates progressively faster and with greater amplitude 
(see Fig.~9(a) for $t<4000\xi/c$).  This leads to an initial {\it decrease} in the oscillation period (Fig.~9(b)) \cite{Brazhnyi}, and 
a significant deviation from the
corresponding energy-conserving trajectory (dashed 
line in Fig.~9(a)).  
However, as the 
soliton begins to probe the outer regions of the trap, where 
the potential gradient tails off, the period {\it increases}. Eventually, at $t\approx6500$ the soliton has lost sufficient 
energy to escape the trap \cite{Parker,Busch,Fedichev}.

The periodic deformation of the soliton as it oscillates in the trap is shown in Fig.~9(c).  The deformation is clearly zero when the
soliton passes 
through the minimum of the trap, where the background density is effectively homogeneous, and becomes maximum at the extremum of each oscillation.  
The energy of the soliton (Fig.~9(d)) decreases in periodic steps, due to the periodic sound emission shown in Fig.~9(e), with both of 
these (shown by solid lines) determined numerically from Eq.~(4).  The peak periodic values of 
both sound emission rate (Fig.~9(e)) and soliton deformation (Fig.~9(c)) initially increase in time as the soliton probes further and
further into the 
double gaussian trap, where the background density gradient becomes {\it higher}.  Later in the evolution, the peaks begin to decrease in 
amplitude and develop intermediate dips as the soliton probes the outer, essentially harmonic regions of the trap 
where the potential tails off and the density gradient {\it decreases}.   
Also plotted in Fig.~9(e) is the power emission 
predicted by Eq.~(7) (dotted line) with $\alpha = 2$ and a suitably chosen value of $\kappa$, such that the initial peak emission
rates match. The agreement is found to be very good, particularly at early times, when the soliton is localised within the central 
region of the trap. 
However, the non-uniformly increasing background density gradient in the trap under consideration causes a degraded agreement
at later times.  We attribute this 
to the deviation of the coefficient $\kappa$ in Eq.~(7) from the
effectively constant 
value assumed here.

\begin{figure}
\begin{center}
\includegraphics[width=12.5cm]{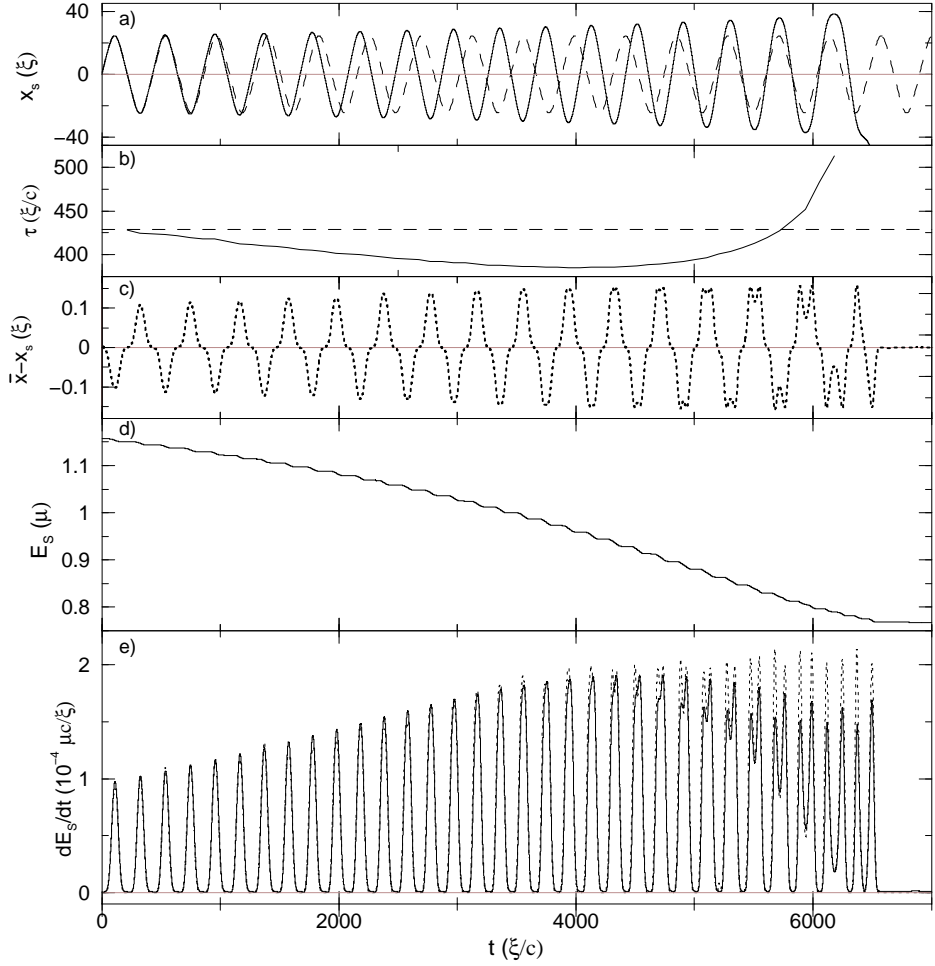}
\end{center}
\caption{a) Path of a dark soliton (solid line) with initial speed $0.3c$ oscillating in the
double gaussian trap of Eq.~(8) with $V_0=0.3 \mu$, $x_0=40\xi$, and $w^2=200\xi^2$, along with the corresponding
non-dissipative trajectory (dashed line). The low amplitude chosen ($V_0=0.3 \mu$) ensures that
the
radiation escapes the inner region.
 b) Oscillation period of the soliton $\tau$ (solid line) and 
effective particle of constant mass (dashed line)
as a function of time 
c) Deformation parameter of the soliton $(\bar{x} - x_{s})$ versus time  d) Decay of soliton energy as a function of time.  e) Power 
emitted by the soliton  as a
function of time computed from the energy functional Eq.~(4) (solid line)
versus the prediction of Eq.~(7) with $\alpha=2$ and $\kappa=0.0107 (\mu c/\xi^3)$ 
(dotted line).} \end{figure}

\begin{figure}
\begin{center}
\includegraphics[width=10cm]{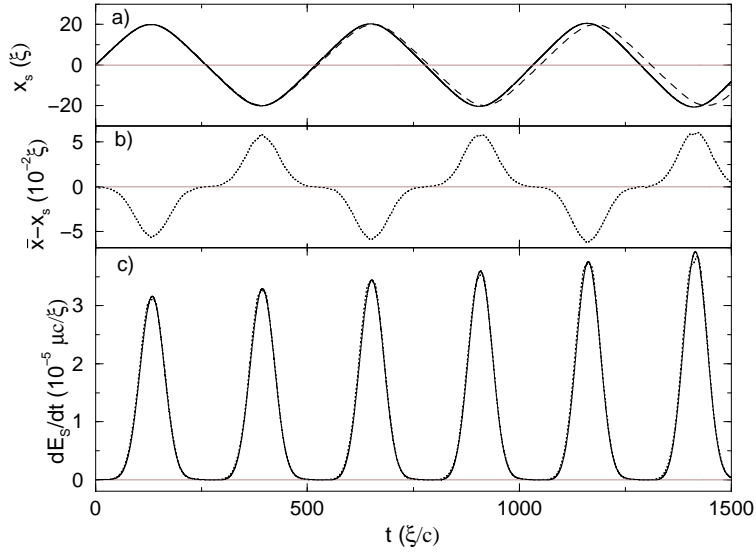}
\end{center}
\caption{a) Path of a dark soliton (solid line) and the corresponding non-dissipative 
particle (dashed line), with initial speed $0.2c$, oscillating 
in 
the same trap of Fig.~9.
b) Deformation parameter of the soliton $(\bar{x} - x_{s})$ as a function of time.  c) Power emitted by the soliton as a
function of time, as computed from the energy functional Eq.~(4) (solid line), 
versus the deformation squared prediction of Eq.~(7) with  
$\alpha=2$ and $\kappa=0.0107(\mu c/\xi^3)$ (dotted line).} \end{figure}

\begin{figure}
\begin{center}
\includegraphics[width=10cm]{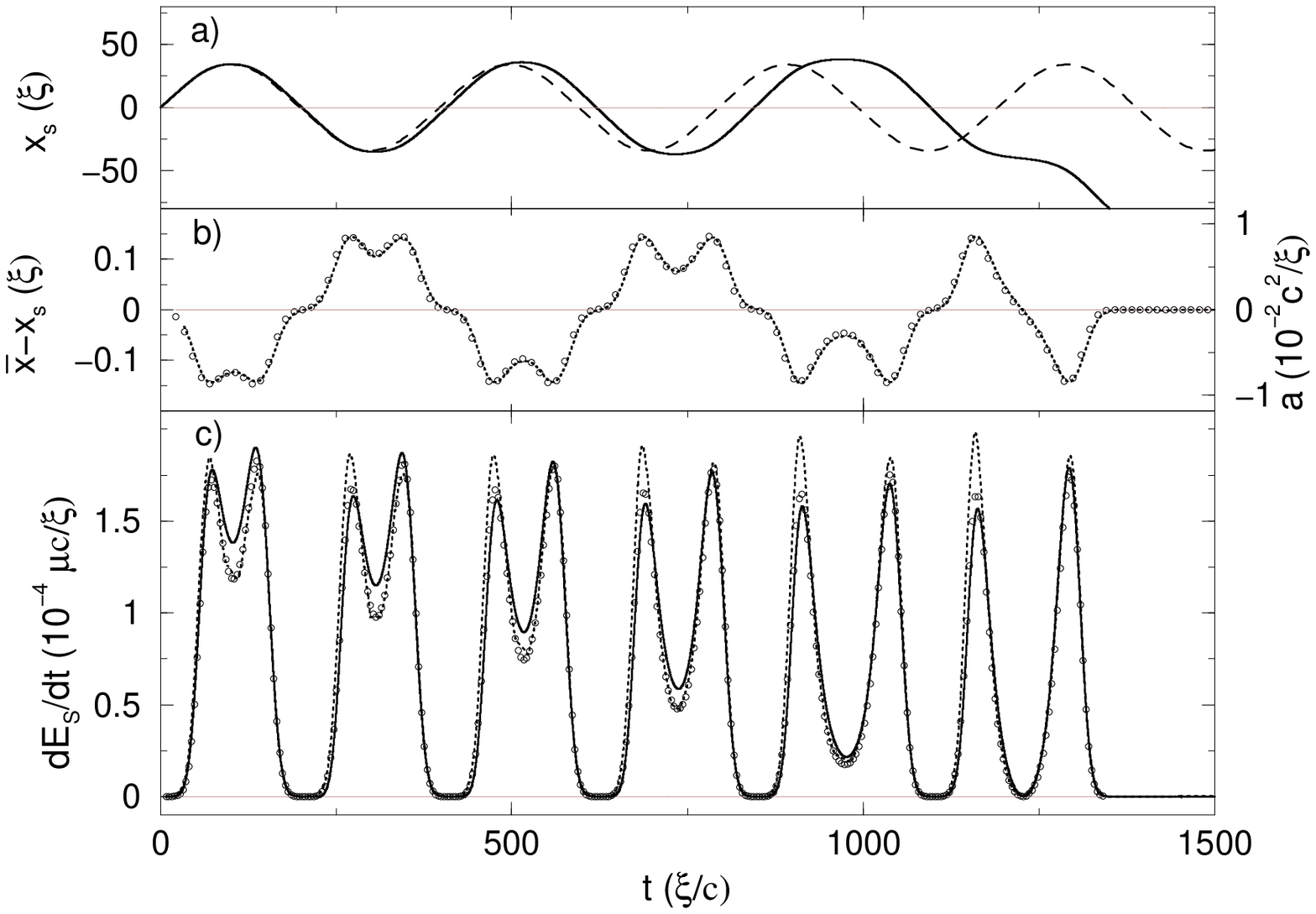}
\end{center}
\caption{a) 
Path of a dark soliton (solid line) with initial speed $0.5c$, and the corresponding non-dissipative 
particle (dashed line), oscillating in the same double gaussian trap as in 
Fig.~9.  b) Deformation parameter of the soliton $(\bar{x} - x_{s})$ 
(dotted line)
(left axis), and soliton acceleration (open circles) (right axis) versus time.  c) Power emitted by the soliton
as a function of time 
computed from the energy functional (Eq.~(4)) (solid line), versus the 
deformation squared prediction of Eq.~(7) with $\alpha=2$ and 
$\kappa=0.0107(\mu c/\xi^3)$ (dotted line).  
Also shown is the acceleration squared power emission predicted by Eq. (9) 
(open circles).} 
\end{figure}

To investigate the validity of Eq.~(7) in more detail, we next investigate the two `limiting' regimes of soliton
motion arising in Fig.~9.
These correspond to (i) the case of a soliton oscillating near the centre of the double gaussian trap (for which the density gradient
increases monotonically as one moves away from the trap centre, roughly $t<2000 (\xi/c)$ in Fig.~9),
and (ii) the limit of a soliton oscillating essentially all the way to the edges of the trap, and thus probing
both the initial monotonic increase of the background density to its maximum value, and its subsequent decrease
(corresponding roughly to $t>4000 (\xi/c)$ in Fig.~9).
To look into these effects more clearly, we investigate respectively the cases of a slower and a faster soliton, compared to the initial
soliton speed of Fig.~9. This will also display more clearly the two competing mechanisms contributing to the change in the
soliton oscillation frequency evident in Fig.~9(a).

Fig.~10 considers a soliton with slow initial speed, $v=0.2c$, oscillating in a double gaussian trap with the same parameters as in Fig.~9.
In particular, Fig.~10(a) shows the deviation of the soliton oscillations from
 the corresponding non-dissipative predictions, with the soliton oscillating faster than the 
non-dissipative case, as a result of sound emission. The resulting soliton deformation is illustrated in Fig.~10(b). Finally, Fig.~10(c)
compares the measured energy loss from Eq.~(4) (solid line) to the prediction of Eq.~(7) (dotted line),
yielding remarkable agreement. 

The opposite regime of a sufficiently fast soliton probing the entire structure of the double gaussian trap is shown in Fig.~11.
In this regime, the soliton experiences more complicated 
dynamics due to the tailing off of the gaussian potential.
Firstly we see that, contrary to the case of a slow soliton (Fig.~10(a)), the fast soliton (Fig.~11(a)) oscillates {\em slower} than the
corresponding non-dissipative trajectory. 
Furthermore, the deformation parameter $(\bar{x} - x_{s})$ plotted in Fig.~11(b) now
develops intermediate dips at the extrema of each oscillation. Both of 
these effects are due to the nature of the changing background density
in the double gaussian trap. Although this gradient initially increases 
with increasing deviation from the trap centre, it
subsequently reaches a maximum and then starts decreasing again, such 
that the soliton period increases drastically as it spends more and more 
time at the edges of the trap. This also explains the intermediate dips in 
the rate of sound
emission shown in Fig.~11(c). Fig.~11(b) further compares the deformation parameter to the local soliton acceleration
computed numerically. This yields an exact proportionality, confirming our earlier claim of Sec.~3. Fig.~11(c) compares the
rate of sound emission in time as computed from Eq.~(4) (solid line), to the prediction of Eq.~(7) based on a {\it constant}
coefficient $\kappa$ (dotted line). Although the intermediate dip is correctly accounted for by Eq.~(7), the incorrect 
assumption of a constant $\kappa$ leads to a slight overestimate of the peak energy emission, with the discrepancy
increasing as the soliton probes the very edge of the trap. These two results are further compared to an 
analytical prediction for sound emission from an unstable dark soliton (on a {\it homogeneous} background) 
derived within the context of nonlinear optics \cite{Pelinovsky}.
The applicability of this theory to inhomogeneous systems has been discussed in \cite{Parker}.
This theory predicts that the breakdown of integrability of the cubic NLSE leads to a rate of energy loss
\begin{eqnarray}
\frac{{\rm d}E_{\rm s}}{{\rm d}t}=-L_{\rm s}(v,n)\left(\frac{{\rm d}v}{{\rm d}t}\right)^2,\label{eqn:kiv}
\end{eqnarray}
where
\begin{eqnarray}
L_{\rm s}(v,n)=\frac{c}{c^2-v^2}&~&\left[\frac{2c^2}{n}\left(\frac{\partial N_{\rm s}}{\partial
v}\right)^2\right.
\\
\nonumber
&~&\left.
+2v\left(\frac{\partial
N_{\rm s}}{\partial v}\right) \left(\frac{\partial S_{\rm s}}{\partial
v}\right)
+\frac{n}{2}\left(\frac{\partial S_{\rm s}}{\partial
v}\right)^2 \right].
\end{eqnarray}
Here $n$ is the local background density, $S_{\rm s}$ the total phase slip across the moving soliton, and
$N_{\rm s}=\int\left(n-|\psi|^2\right){\rm{d}}z$ is the number of particles displaced by the soliton.

The prediction of Eqs.~(9)-(10) shown by the open circles in Fig.~11(c) indeed yields near perfect agreement with the
energy functional calculations of Eq.~(4).  The peaks in power emission are now correctly accounted for, due 
to the
constantly adjusting proportionality coefficient $L_{\rm s}(v,n)$ of 
Eq.~(9). Since Fig.~11(b) proves an exact
proportionality between acceleration and deformation parameter, we conclude that the inadequacy of Eq.~(7) in the
limit of solitons probing the edge of the trap is due to the oversimplified assumption of a constant
coefficient of proportionality $\kappa$. Assigning a slightly varying $\kappa$ which depends on the local
$n$ and $v$ in a manner similar to the dependence of $L_{\rm s}$ on these parameters will clearly remedy
this apparent inaccuracy of Eq.~(7). Nonetheless, it is remarkable that even the oversimplified assumption
of a {\it constant} $\kappa$ yields reasonable values for the rate of sound emission.

\section{Discussion}

We have examined soliton dynamics and sound emission
due to various forms of background density perturbations in one dimension
and have quantified the sound emission in terms of the step geometry.
We have shown that a dark 
soliton incident upon a potential step can emit a significant fraction of its energy.
For example, a soliton with initial speed $v=0.7c$ can emit  {\it more than $60\%$ of its energy}
in its interaction with the step.

By considering smoother density inhomogeneities, we have shown that the gradient of the background density 
results
in a deformation of the soliton profile,
in the sense that the soliton centre of mass on an inhomogeneous
background deviates from the soliton density minimum.
The reason for this `mismatch' is clear: As the soliton propagates
through the inhomogeneous region, it is induced to emit counter-propagating waves
(with positive and negative amplitudes with respect to the background density),
each of which moves with the local speed of sound.
However, the {\em relative} motion of the emitted sound waves to the
soliton minimum depends on the direction of motion of the soliton.
This means that the sound wave moving in the soliton direction 
with speed $(c-v)$ is more
compressed than the one moving in the opposite direction at speed ($c+v$). 
We note that, strictly speaking, one cannot distinguish between soliton and sound energy
trapped within the soliton region, and this would also be true for any
experimental measurement of the soliton profile.
Hence, the soliton profile essentially includes the amount of emitted sound which
has not yet escaped the soliton region and this leads to
the apparent deformation of the soliton.
This deformation was shown to be directly proportional to the acceleration of the soliton through the 
fluid (determined as the acceleration of the density minimum). 
More importantly we have shown that the energy emitted by the soliton due to
background inhomogeneities can be parametrised in terms of the square of both this deformation
parameter and, hence, of the acceleration. The coefficient of proportionality
is, to reasonable approximation constant, depending on initial soliton speed and 
background density gradient. However, a more accurate determination would require
the calculation of its value at each position,
since, in general, it will depend on the
{\it local} soliton speed and background
density.

\ack
We acknowledge financial support from the UK EPSRC.


\begin{thebibliography}{99}
\bibitem{comment} For a harmonic trap, soliton deformation has also been found in \cite{Brazhnyi}.
\end{thebibliography}
\end{document}